\documentclass{appolb}
\usepackage{graphicx}

\usepackage[T1]{fontenc}
\usepackage[utf8]{inputenc}  
\usepackage{amsmath}
\usepackage{cite}     
\usepackage{upgreek}  
\usepackage{hyperref} 
\usepackage{placeins}

\begin{document}
\title{J-PET: a new technology for the whole-body PET imaging%
\thanks{Presented at the 2\textsuperscript{nd} Jagiellonian Symposium on Fundamental and Applied Subatomic Physics, Krak\'ow, Poland, June 4--9, 2017. }%
}

\author{
	S.~Nied\'zwiecki$^a$,   
	P.~Bia\l{}as$^a$, C.~Curceanu$^b$, E.~Czerwi\'nski$^a$, K.~Dulski$^a$, A.~Gajos$^a$, B.~G\l{}owacz$^a$, M.~Gorgol$^c$, B.~C.~Hiesmayr$^d$, B.~Jasi\'nska$^c$, \L. Kap\l on$^{a}$
D.~Kisielewska-Kami\'nska$^a$, G.~Korcyl$^a$, P.~Kowalski$^{e}$, T.~Kozik$^a$, N.~Krawczyk$^a$, W.~Krzemie\'n$^{f}$, E.~Kubicz$^a$, M.~Mohammed$^{a,g}$, 
M.~Pawlik-Nied\'zwiecka$^a$, M.~Pa\l{}ka$^a$, L.~Raczy\'nski$^{e}$, Z.~Rudy$^a$, N.~G.~Sharma$^a$, S.~Sharma$^a$, R. Y.~Shopa$^{e}$, M.~Silarski$^a$, M.~Skurzok$^a$, 
A.~Wieczorek$^a$, W.~Wi\'slicki$^{e}$, B.~Zgardzi\'nska$^c$, M.~Zieli\'nski$^a$, P.~Moskal$^a$
\address{
$^{a}$ Faculty of Physics, Astronomy and Applied Computer Science, Jagiellonian University, 30-348 Cracow, Poland \\
$^{b}$ INFN, Laboratori Nazionali di Frascati, 00044 Frascati, Italy \\ 
$^{c}$ Institute of Physics, Maria Curie-Sk\l{}odowska University, 20-031 Lublin, Poland \\
$^{d}$ Faculty of Physics, University of Vienna, 1090 Vienna, Austria \\
$^{e}$ Department of Complex Systems, National Centre for Nuclear Research, 05-400 Otwock-\'Swierk, Poland \\
$^{f}$ High Energy Physics Division, National Centre for Nuclear Research, 05-400 Otwock-\'Swierk, Poland \\
$^{g}$ Department of Physics, College of Education for Pure Sciences, University of Mosul, Mosul, Iraq
}
}

\maketitle
\begin{abstract}
The  Jagiellonian Positron Emission Tomograph (J-PET) is the first PET built from plastic scintillators. J-PET prototype consists of 192 detection modules arranged axially in three layers forming a cylindrical diagnostic  chamber with the inner diameter of 85 cm and  the axial  field-of-view  of  50  cm.  An axial arrangement of long strips of plastic scintillators, their small light attenuation, superior timing properties, and relative  ease  of  the increase of the axial field-of-view opens promising perspectives for the cost effective construction of the whole-body PET scanner, as well as construction of MR and CT compatible PET inserts. Present status of the development of the J-PET tomograph  will be presented and discussed.
\end{abstract}
\PACS{29.40.Mc, 87.57.uk, 87.10.Rt, 34.50.-s}
  
\section{Introduction}
Positron emission tomography is a medical technique used mainly for cancer studies as well as control of radio- and chemo-therapies. Before examination, the patient is being injected with a radioactive marker, which emits positrons. After traveling for short distance, positron-electron annihilation occurs and in most cases a pair of almost back-to-back $\gamma$ quanta are produced. 
Reconstruction of the annihilation position, by using informations from annihilation quanta, is providing a spatial density distribution of injected marker inside patients body. By selecting different tracers, one can select different metabolical processes to observe during a scan.
All available state of the art scanners are detecting $\gamma$ quanta by crystal scintillators \cite{Karp2008, Slomka2016, EJNMMI2016}. Their main advantage is large stopping power, high probability of photoelectric effect and good energy resolution. 
\newline
\indent 
One of the challanges in the PET tomography is the simultaneous imaging of the whole human body.
Due to the high cost of crystal scintillators, a production of the commonly available whole body scanner based on crystals seems implausible. Currently only about 20 cm along the body can be simultanousely examined at single bed position \cite{Karp2008}. In case of whole-body scan, several overlapping bed positions are needed.
 Currently only ~1\% of $\gamma$ quanta emitted from patien's body are collected \cite{Cherry2017}. Extension of the scanned
 part from around 20 cm to 200 cm would improve the sensitivity and signal-to-noise ratio. The radiation dose needed for whole body scan can be also reduced and usage of shorter living tracers will be simplified. 
To address this problem several different designs of whole body scanners were introduced based on resistive plate
chamber (RPCs) \cite{Blanco2006}, straw tubes \cite{Sun2007, Shehad2005} and crystal scintillators \cite{Cherry2017}.
\newline \indent
The J-PET group proposes the usage of plastic scintillators as detection material for positron emission tomography \cite{Moskal2011}. This will allow a
construction of cost effective whole-body scanner, due to less expensive detector material \cite{Moskal2011, Moskal2014}. In
addition the readout can be placed outside of detection chamber simplifying PET/MR hybrid construction and enabling extension of the axial field of view (AFOV) without significant increase of costs. The costs of the electronic readout is not changing for the extended J-PET, because the number of photomultipliers and electronic channels remains independent of the AFOV.
In order to compare a performance of the crystal based PET tomographs and the J-PET built from strips of plastic scintillators we introduced a figure of merit (FOM) for the whole body imaging \cite{MoskalRundel2016} by analogy to the figure of merits proposed earlier in \cite{Conti2009, EiC2015}. 
\newline \indent
The whole body FOM is defined as a 
probability of detection of annihilation event divided by the Coincidence Resolving Time (CRT) and the number of bed positions. 
Comparions of such introduced FOM for the J-PET and scanner based on LSO crystals, with AFOV = 20 cm and CRT = 400 ps shows that one can overcome lower probability of detection of plastic
scintillators by using longer modules and more detection layers. For 50 cm long plastic modules, one expects already the
same performance for a whole-body scan as for commercial scanners, while the introduction of second layer should improve it a few times. It is
worth to note that the plastic scintillator modules could be even 2 m long, but this comes with a trade off the CRT,
which will decrease with elongation of modules \cite{MoskalRundel2016}.
\newline
\indent
In this paper the general concept of the J-PET scanner is described. Then, the previous prototype built out of 24
modules is presented along with the latest full scale prototype, based on 192 detection modules with 50 cm AFOV and
85 cm diameter of first layer and 115 cm diamater of last one. Initial TOF resolution studies are presented. Finally, conclusions and perspectives are given.
  
\section{General concept of J-PET scanner}

\begin{figure}[h]
\begin{center}
\includegraphics[width = 0.4\textwidth]{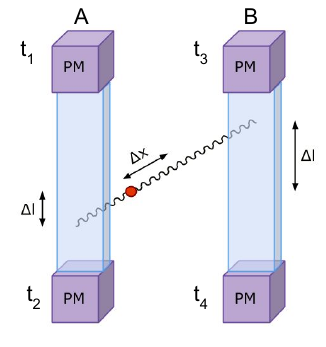}
\end{center}
\caption{Reconstruction of annihilation point can be determined by the usage of times denoted as $t_i$.
\label{fig:recoConcept}
}
\end{figure}

The J-PET tomgraph is constructed from axially arranged strips of plastic scintillators. Annihilation $\gamma$ quanta with energy of 511 keV interact in plastic scintillators through Compton effect, in which deposited energy varies from event to event.
Due to the low light attenuation, plastic scintillators act as effective lightguides for photons produced by interaction of radiation, hence examination chamber can be built out of long modules placed along patients body.
Each plastic strip is read out by photomultipliers at two ends (Fig.~\ref{fig:recoConcept}).
Since the readout will be placed outside of the diagnostic chamber, the main cost of extending the axial field of view of the scanner lays in cost of scintillating material.  
\newline \indent
The position of interaction of $\gamma$ quanta with scintillator can be determined from time difference of light signal arriving at photomultipliers placed at each end of detection module:
\begin{equation}
\Delta l_A = (t_1 - t_2)* V_A,
\end{equation}
where $\Delta l_A$ denotes the distance between the interaction point and the center of module, $t_1$ and $t_2$ stand
for times of arrival of light signal at each side of scintillator and $V_A$ is an effective velocity of light signal within the scintillator.
Then, the position of annihilation along Line Of Response (LOR) can be determined by usage of TOF method (see Fig.~\ref{fig:recoConcept} for pictorial description):
\begin{equation}
TOF =  (t_1 + t_2)/2 - (t_3 +t_4)/2; \hspace{1cm} 
\Delta x = TOF * c/2;
\end{equation}
where $\Delta x$ denotes distance of annihilation point from middle of LOR, and $c$ stands for the speed of light.

\begin{figure}[h]
\begin{center}
\includegraphics[width = 0.8\textwidth]{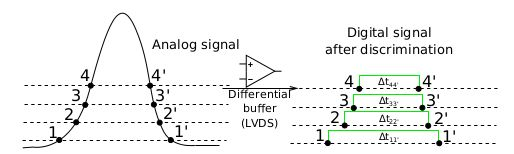}
\end{center}
\caption{Pictorial representation of signal probing. After signal proccessing four pairs of points are acquired at four selected thresholds. This figure is adapted from \cite{Palka2017}.
\label{fig:signalProbing}
}
\end{figure}

The J-PET predominantly uses time information instead of energy to acquire place of annihilation. 
Scintillating signals from plastics are very "fast" (typically 0.5 ns rise time and 1.8 ns decay time \cite{Wieczorek2017}). Such fast signals allow for superior time resolution and decrease pile-ups with respect to crystals detectors as e.g. LSO or BGO with decay times equal to 40 ns and 300 ns, respectively \cite{Saha}.
In order to take advantage of this superior timing properties of plastic scintillators and to decrease the dead time due to the electronic signal processing in J-PET, the charge measurement was replaced with measurement of TOT (Time Over Threshold). 
In order to improve the resolution of the scanner, signals are probed at four different constant thresholds at the rising and falling edge. The sampling of the signal depicted schematically in Fig. \ref{fig:signalProbing} is performed by the newly developed method based solely on FPGA units \cite{Palka2017}. 
Sampling at few selected thresholds gives opportunity to improve the resolution of determination of time and place of
interaction of $\gamma$ quanta with detector by applying more advanced methods of reconstruction published in \cite{Raczynski2017, Raczynski2015, Raczynski2014, MoskalSharma2015, Sharma2015}.
Data acquisition is performed in trigger-less mode and can handle stream of data up to 8 Gbps \cite{Korcyl2016}.
Utilisation of digital time measurement decreases price of the electronics as well as power consumption. 

\section{The 24th-modules  prototype}
\begin{figure}[h]
\label{fig:smallBarrel}
\begin{center}
\includegraphics[width = 0.7\textwidth]{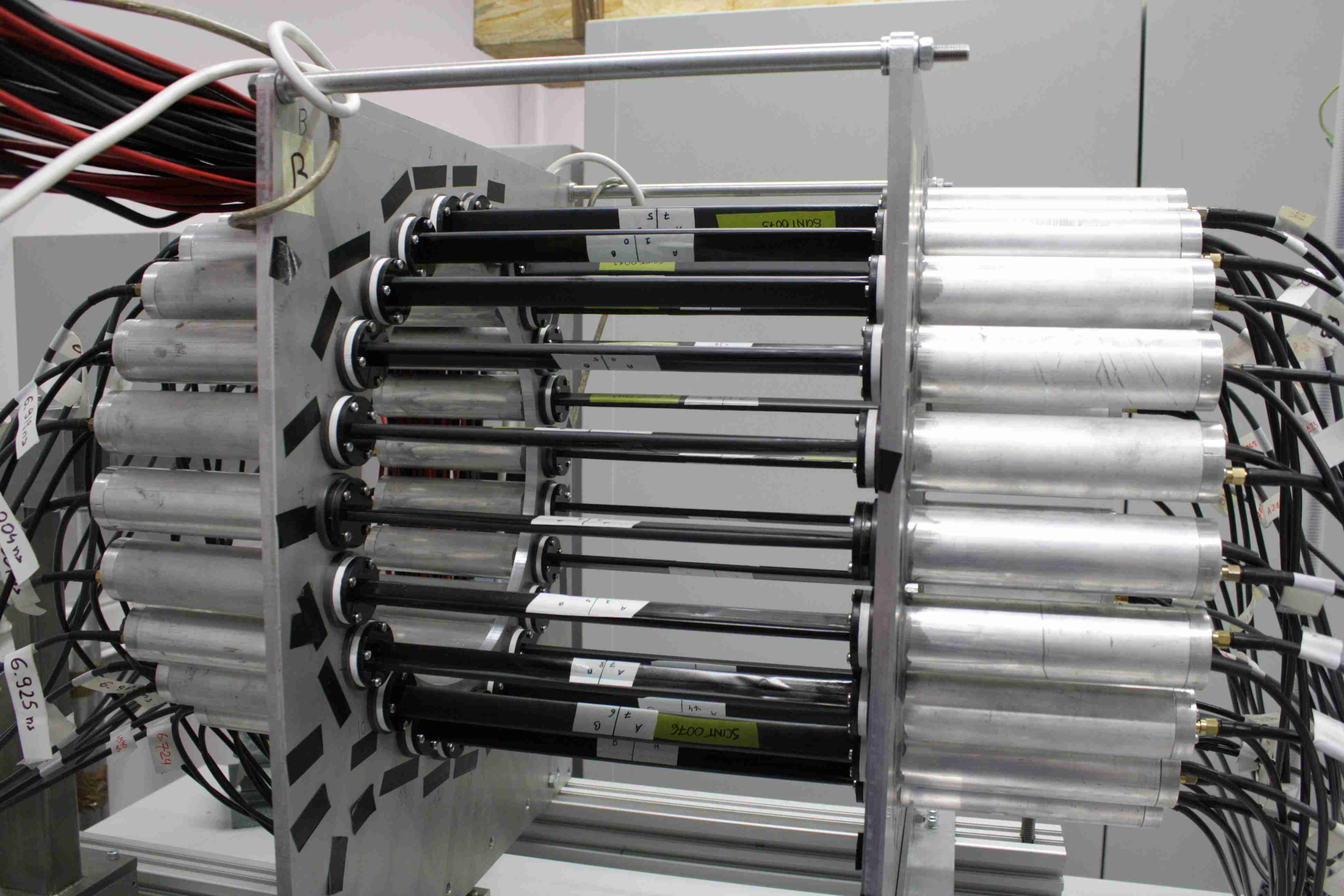}
\end{center}
\caption{24th-modules prototype of J-PET scanner. BC-420 scintillators (covered with black foil) were forming single ring with 360 mm diameter. To each side of scintillator a R4998 Hamamatsu photomultipliers were connected.}
\end{figure}

The first working prototype of J-PET scanner, shown in Fig.~\ref{fig:smallBarrel}, was constructed from 5x19x300 mm$^3$ strips of BC-420 scintillators formed in a ring with 360 mm diameter. Each scintillator was read out by R4998 Hamamatsu photomultipliers  and signals were probed at four levels by front end boards \cite{Palka2017}. 
The main reason to build this prototype was to test electronic readout, develop calibration procedures and proceed from two modules studies to system where one has to control more detectors. The resulting CRT for this prototype was equal to 490 ps \cite{pracaTB} which is comparable to the best currently available scanners \cite{Karp2008, Slomka2016, EJNMMI2016}. Unfortunately, the amount of detection modules was not enough for the effective image reconstruction inside the chamber due to too many holes in acceptance. 
Nonetheless we have gained the necessary experience to know how to design and construct the full scale prototype.

\section{First full scale J-PET prototype} 
\begin{figure}[h]
\begin{center}
\includegraphics[width = 0.58\textwidth]{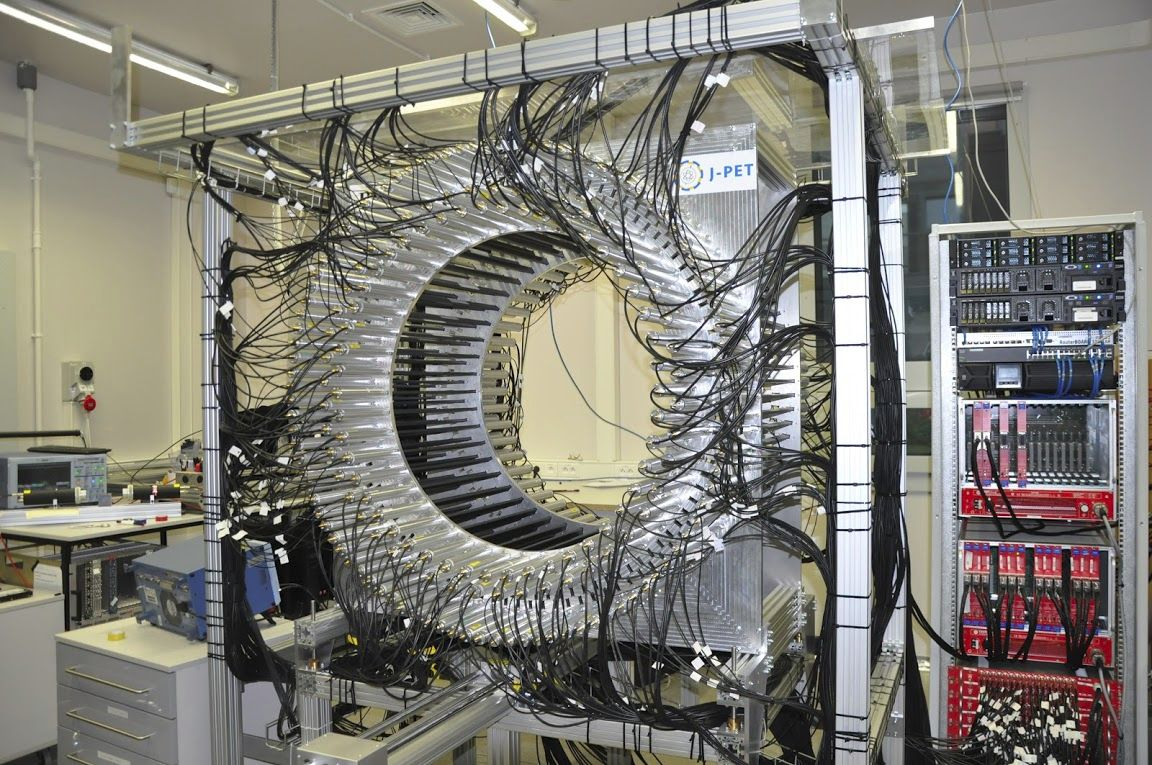} 
\includegraphics[width = 0.40\textwidth]{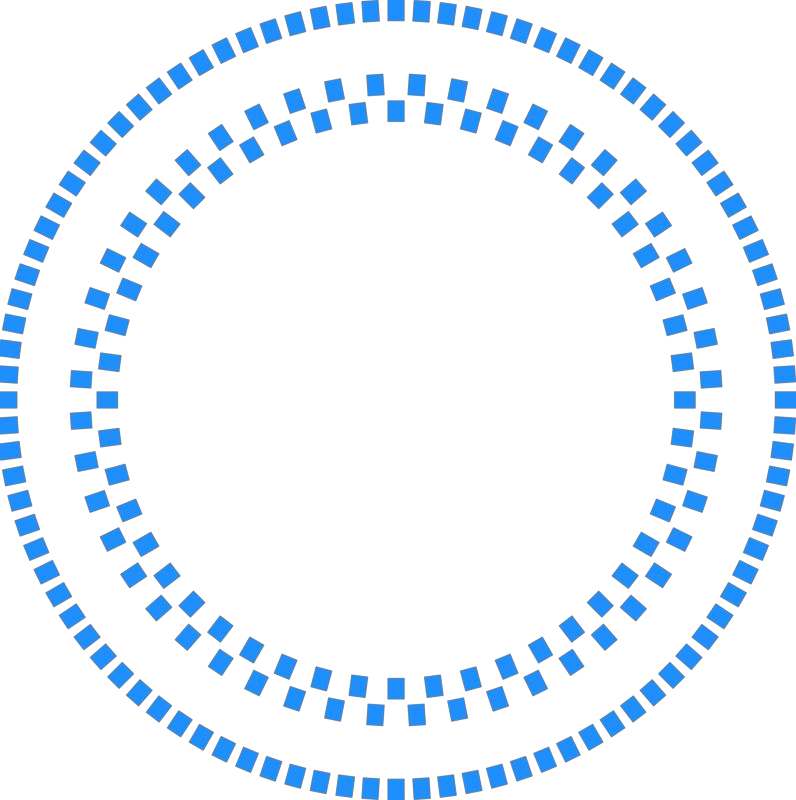}
\end{center}
\label{fig:bigBarrel}
\caption{Photo of the full scale J-PET prototype (left) and layout of detection rings (right). Detection modules are arranged into three non-overalaying rings which cross section is shown schematically on the right panel. Diameters and amount of detection modules from the most internal layer are following: 850 mm and 48, 935 mm and 48, 1150 mm and 96.}
\end{figure}

A full scale version of the prototype has been assembled from 192 detection modules, with each module constructed from
7x19x500 mm$^3$ EJ-230 scintillators with two R9800 Hamamatsu photomultipliers coupled to each end. 
Modules were arranged in three, not overlaying rings (see Fig. \ref{fig:bigBarrel}). As with previous prototype, this setup was read-out by multi-constant-threshold boards \cite{Palka2017} combined with TRB3 boards \cite{Traxler2004}. 
At  present the prototype is in a commissioning and calibration phase. 

\begin{figure}[h]
\begin{center}
\includegraphics[scale = 0.3, width=0.39\textwidth]{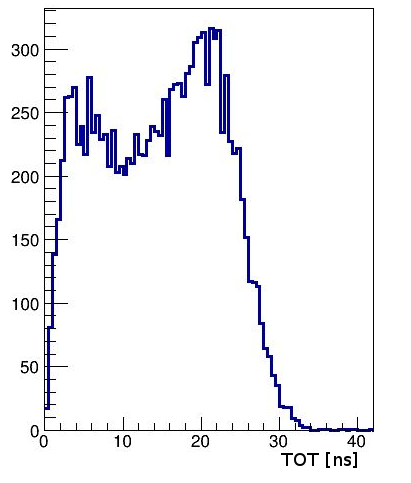} 
\end{center}
\caption{Distribution of the TOT sum from four thresholds at two photomultipliers connected to the same scintillator. Only back-to-back $
\gamma$ quanta were selected. Test setup consisted of $^{22}$Na source placed inside lead collimator \cite{Kubicz2016}. Source was placed at the center of detection chamber.
\label{fig:TOTsum}
} 
\end{figure}

\begin{figure}[h]
\begin{center}
\includegraphics[width = 0.80\textwidth]{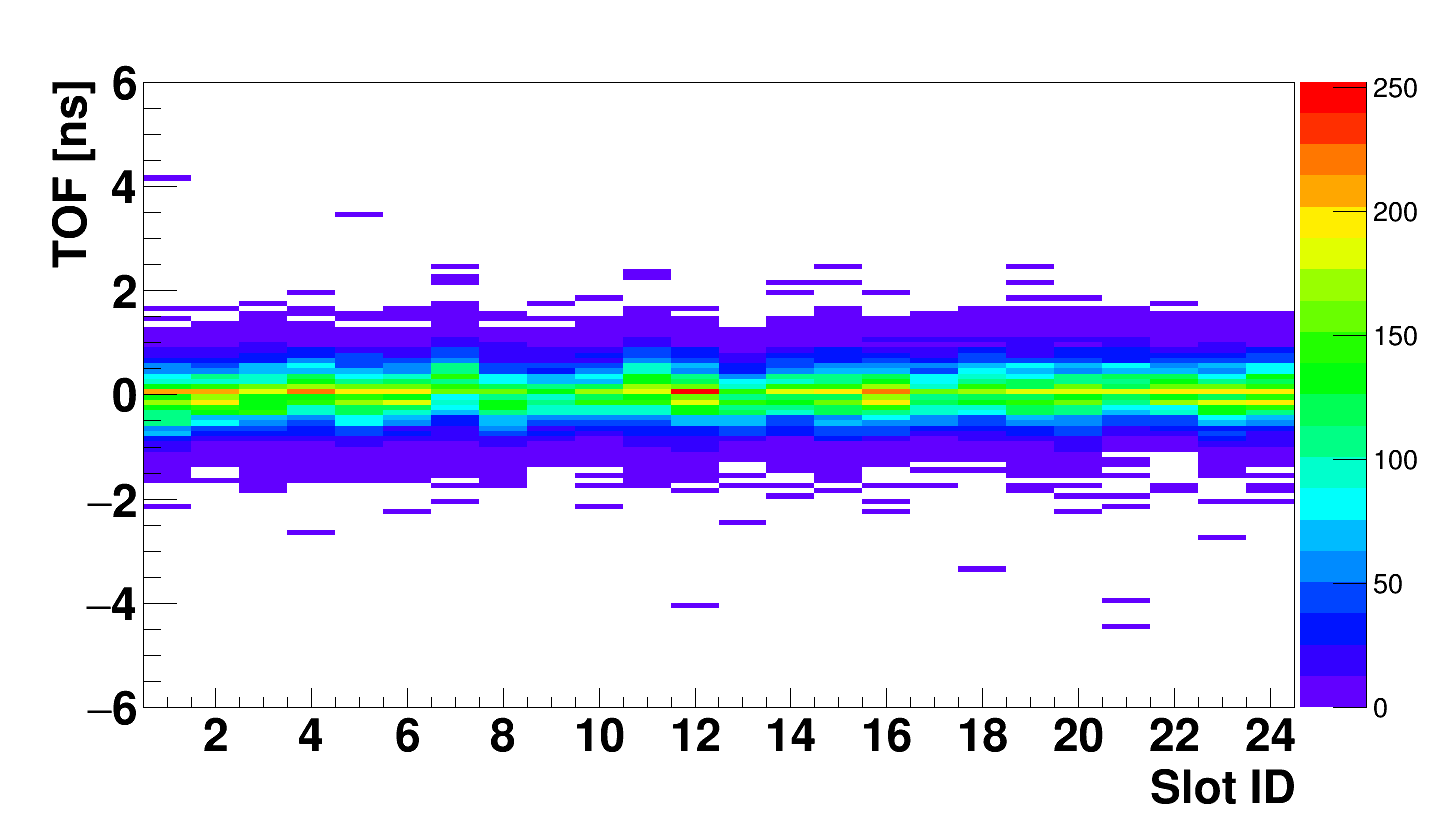}
\end{center}
\caption{Histograms of TOF, measured by scintillators from the second layer modules after calibration \cite{Skurzok2017}, versus scintillator ID in the second layer. Test setup consisted of $^{22}$Na source placed inside the lead collimator. The collimator was constructed from two lead disks mounted on long arms extending outside of detection chamber  \cite{Kubicz2016}. Source was placed at the center of detection chamber. Only back-to-back events were taken into account. Second layer includes 48 scintillators, thus only 24 back-to-back pairs are presented.
\label{fig:tofCalib}
}
\end{figure}

\begin{figure}[h]
\begin{center}
\includegraphics[width = 0.48\textwidth]{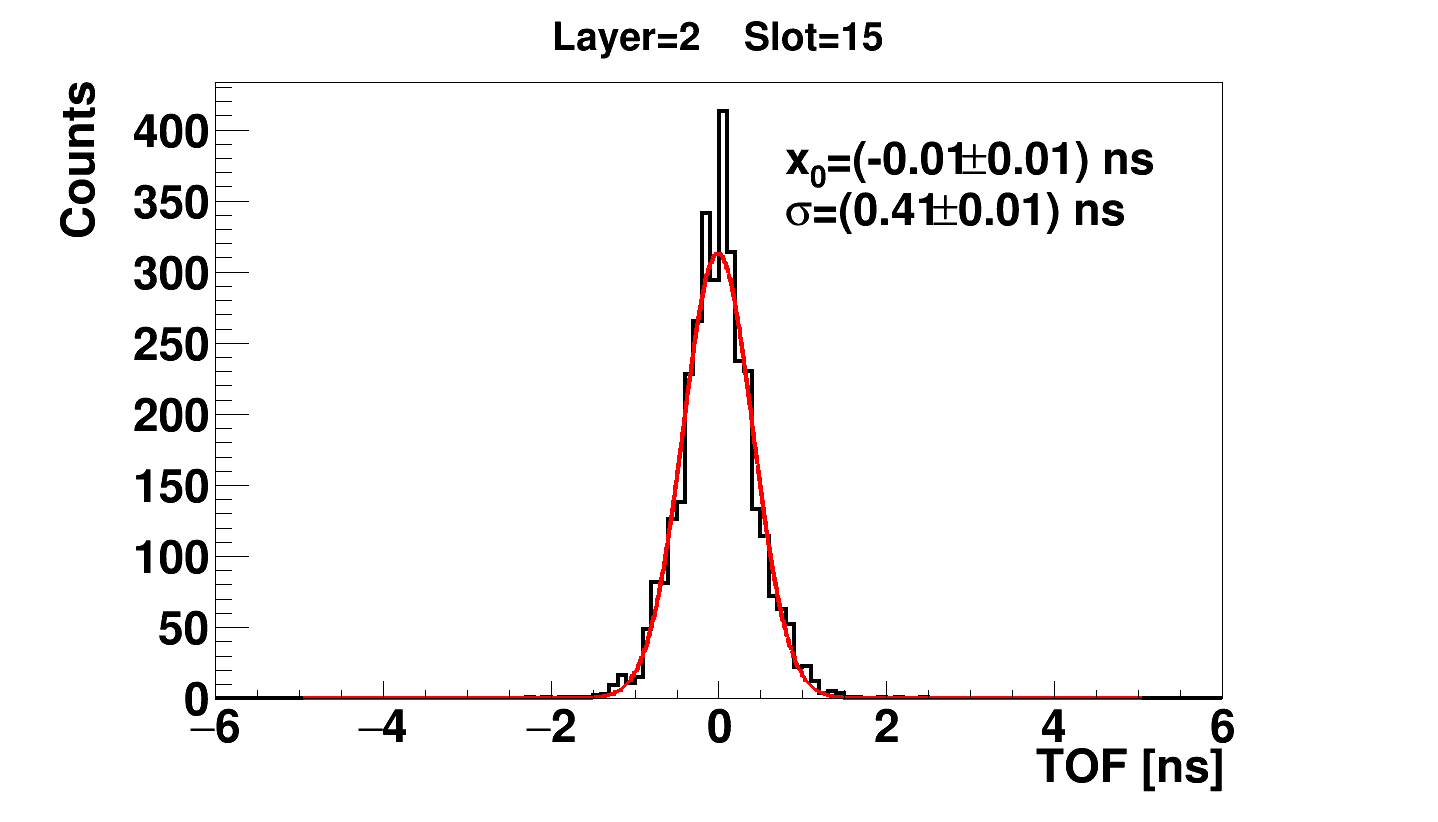}
\includegraphics[width = 0.48\textwidth]{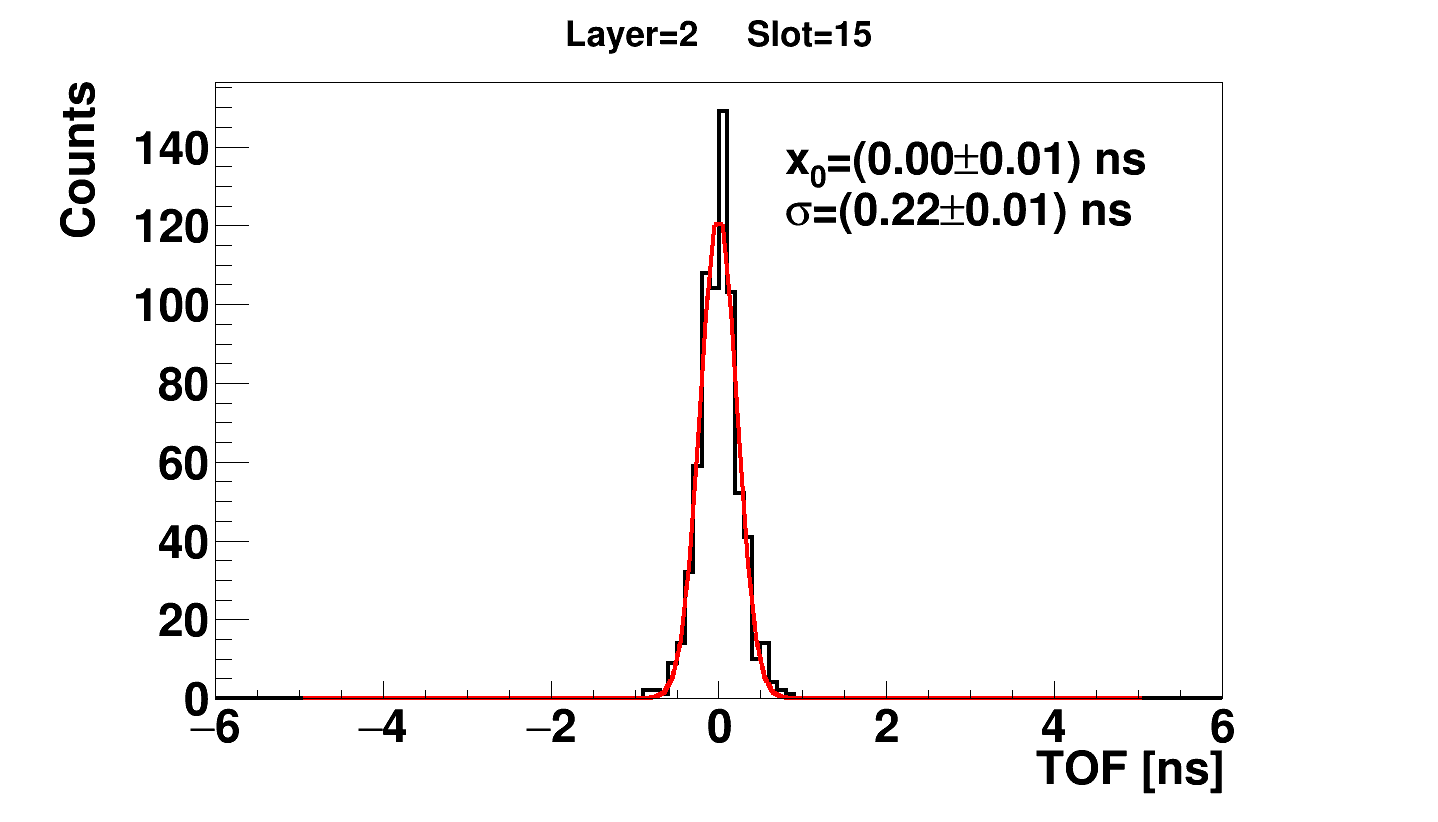}
\end{center}
\caption{TOF spectrum for one module from the second layer before- (left) and after (right) the cut on TOT spectrum. $^{22}$Na source was placed inside lead collimator in the center of detection chamber. TOT cut was set to value corresponding to 200 keV of deposited energy.
\label{fig:TOFTOTcuts}
}
\end{figure}

\indent In a first approximation, as a measure of the energy deposition a sum of Time Over Threshold (TOT) values was used:
\begin{equation}
TOT = \sum_{side = A,B} \sum_{thr = 1}^{4} TOT_{side,thr},
\end{equation}
\noindent where, $A,B$ denote left and right photomultipier and subscript $thr$ denotes the number of the selected threshold. 
An exemple sum of TOT spectrum is presented in Fig. \ref{fig:TOTsum}. 
Since maximum energy deposited by annihilation $\gamma$ quanta in plastic scintillators is equal to about 
340 keV, an energy threshold of 200 keV will reduce the scattering of gamma quanta in the body of a patient to about the same level \cite{Kowalski2016} as in previous commercial tomographs with low energy threshold of 300 or 350 keV \cite{Humm2003}. 
Fig.~\ref{fig:tofCalib} shows an exemplary spectrum of time of flight  distributions as a function of the scintillator identifier (ID)  after the first iteration of time synchronisation. 
In Fig.~\ref{fig:TOFTOTcuts} an exemplary TOF distributions are presented before and after cut on Time Over Threshold
corresponding to 200 keV.
The results presented in aforementioned Figs. were obtained using J-PET Analysis Framework
\cite{Krzemien}.
\FloatBarrier
\section{Conclusions and perspectives}
In this article, first preliminary results from the commissioning of the first tomograph built from plastic
scintillators were presented. Whole body J-PET prototype consists of 192 detection modules arranged axially in three layers forming a cylindrical diagnostic  chamber with the inner diameter of 85 cm and  the axial  field-of-view  of  50  cm.  
In order to take advantage of fast scintillating signals a Time Over Threshold is used to determine energy deposition inside detection modules. 
Due to selection of the material, such scanner will constitute cost-effective solution to whole body scans. The placement of detection modules along the patient's body and readout at their end will reduce the price for the extention of the detection chamber, since main expeditures will be spent on plastic scintillators. Such design will also simplify the construction of PET/MR hybrids due to the placement of photomultipliers beyond magnetic fields. 
Preliminary studies of Coincidence Resolving Time show that it is possible to achieve the resolution of 220 ps ($\sigma$).
One should note, that this result is obtained by using only a single threshold in time measurement and it can be improved by the utilisation of more complex methods.
Three approaches were already checked: compressive sensing theory \cite{Raczynski2015, Raczynski2016Nuk}, comparison of acquisited signal with averaged signals \cite{MoskalSharma2015} and with library of model signals \cite{MoskalZon2015}. 
Initial Point Spread Function studies reported in ref. \cite{Shopa2017} show that values equal to 5-7 mm and 9-20 mm for
transverse and longitudinal directions can be obtained. In order to improve the resolution along scanner length a method
utilising Wavelength Shifting strips placed perpendicular to the detection modules was tested \cite{Smyrski2017},
indicating that even 5 mm ($\sigma$) resolution can be achieved, with not optimized setup, leaving room for further improvement by optimisation
of the WLS parameters and plastic strips.
As next steps we intend to improve data selection and estimation of time of the interaction with the scintillator.

\section*{Acknowledgements}
The authors acknowledge the technical support by A. Heczko, W. Migda\l \newline  and the financial support from the Polish National Center for Research and Development through grants INNOTECH-K1/IN1/64/159174/NCBR/12, and the LIDER-274/L-6/14/NCBR/2015, the EU and MSHE Grant no. POIG.02.03.00-161 00-013/09, National Science Center Poland through grant No. 2016/21/B/ST2/01222 and by the Ministry for Science and Higher Education through grants No. 7150/E-338/M/2017 and 7150/E-338/SPUB/2017/1. B. C. Hiesmayr acknowledges
gratefully the Austrian Science Fund FWF-P26783.


\end{document}